\documentclass[12pt]{article}
\usepackage{epsfig}
\begin{document}
\begin{titlepage}
\begin{center}

{\Large Non-extensive statistical mechanics and particle spectra in 
elementary interactions} 

\vspace{2.cm}
{\bf Christian Beck}\footnote{
permanent address:
School of Mathematical Sciences, Queen Mary and Westfield College,
University of London, Mile End Road, London E1 4NS, England}

\vspace{1cm}
Institute for Theoretical Physics, University of California at
Santa Barbara, California 93106-4030, USA

\end{center}

\vspace{3cm}

\abstract{We generalize Hagedorn's statistical theory of momentum spectra
of particles produced 
in high-energy collisions using Tsallis' formalism
of non-extensive statistical mechanics. Suitable non-extensive 
grand canonical partition functions are introduced for both fermions and
bosons. Average occupation numbers and 
moments 
of transverse momenta are evaluated in an analytic way. 
We analyse the energy dependence of the non-extensitivity parameter $q$
as well as the $q$-dependence of the Hagedorn temperature. We
also take into account the multiplicity. As a final result we
obtain  
formulas for differential cross sections that are in  very good
agreement with $e^+e^-$ annihilation experiments.} 

\end{titlepage}

\section{Introduction}

35 years ago Hagedorn has written a paper of fundamental
importance \cite{hage} which nowadays is very often cited in various
contexts of statistical mechanics, particle physics, and string theory.
In this paper he developed a statistical
description of momentum spectra of particles produced in collider
experiments. At a fundamental level, of course, the underlying
theory for this
is quantum chromodynamics. However, for the hadronization
cascade  phenomenological models
have to be used in addition to QCD. 
The complexity inherent in this
process is so immense that usually one has to perform Monte Carlo
simulations in order to explain the experimental measurements
of cross sections. 

Hagedorn's successful approach was to see this
problem from a thermodynamic point of view (of course QCD was
not known at the time he wrote the paper). He devoped a theory
that is nowadays regularly in use do describe 
e.g.\ heavy ion collisions \cite{ion1,ion,ion2}. The
basic assumption is that in the scattering region
the density of states grows so rapidly that
the temperature cannot exceed a certain maximum temperature, the
Hagedorn temperature $T_0$. This is similar to a first oder phase
transition of e.g. boiling water, just that the Hagedorn temperature
describes 'boiling' nuclear matter at about $T_0\approx 200$ MeV.
This state of matter is closely related
to what is nowadays called quark gluon plasma.
For some theoretical approaches to transverse
momentum spectra, see e.g.
\cite{theor,theor2,theor3}.
As already mentioned, the Hagedorn phase transition is also
of
fundamental interest 
in string theories \cite{string1, string2, string3}.

While for collider experiments (e.g.
$e^+e^-$ annihilation) the Hagedorn theory yields a good description
for relatively small center of mass enegies ($E< 10$ GeV), it fails
at large energies. Indeed, Hagedorn's approach predicts an exponential
decay of differential cross sections for large transverse momenta,
whereas in experiments one observes non-exponential behaviour for
large energies ($E> 10$ GeV) \cite{TASSO, DELPHI, ZEUS}. 
For this reason one usually restricts
the comparison of the experimental results at higher energies 
to Monte Carlo simulations, which reasonably well reproduce the experimental
data. But apparently, for a deeper understanding
from a statistical mechanics point of view
there is the need to generalize Hagedorn's original ideas.

In this paper we show that it is possible
to extend Hagedorn's theory in such a way that
it correctly describes the experimental findings
at large energies as well.
The basic input for this is
the recently developed
formalism of non-extensive statistical mechanics \cite{1, 2, 3}, 
a generalization of ordinary statistical mechanics suitable for
multifractal and self-similar systems with 
long-range interactions. The formalism has been introduced by Tsallis
12 years ago \cite{1} and has in the mean time been 
shown to be
very successful to describe e.g. systems exhibiting turbulent behaviour
\cite{hydro,  ari, 12} or Hamiltonian 
systems with long-range interactions \cite{6}. 
Several other physical applications were described in 
\cite{7}---\cite{10}. 
Very recently, evidence was provided that the new formalism
has applications in particle physics as well.
Walton and Rafelski \cite{raf} studied a Fokker Planck equation describing
charmed quarks in a thermal quark-gluon plasma and showed that
Tsallis statistics is relevant. 
Wilk and Wlodarczyk \cite{wilk} provided evidence that 
the distribution of depths of vertices of hadrons originating from
cosmic ray cascades, as measured in emulsion chamber
experiments, follows Tsallis statistics.
Alberico, Lavagno and Quarati
\cite{qua} successfully analysed heavy ion collisions using Tsallis statistics.
Bediaga, Curado and Miranda \cite{11} showed that differential cross sections
in $e^+e^-$ annihilation experiments can be very very well fitted
using Tsallis statistics. 
In the present paper we will work out the
theory underlying this.

To effectively describe the complex QCD interactions and
hadronization cascade by a
a thermodynamic model, it seems reasonable
to use the Tsallis formalism
(which contains ordinary statistical mechanics as a special case), since
this formalism is specially designed to include selfsimilar systems and systems
with long-range interactions. What happens in a collider experiment
is expected to fall into this category, since QCD forces are strong at
large distances. Moreover,
the self-similarity of the scattering process was already recognized
by Hagedorn, who described the various possible
particle states as a 'fireball' and who defined a fireball
as follows \cite{hage2}: {\it A fireball is}

\vspace{0.2cm}

{\it *... a statistical equilibrium of an undetermined number of
all kinds of fireballs, each of which in turn is considered to be...}

(back to *)

\vspace{0.2cm}

Clearly, nowadays we would call this a self-similarity
assumption.

In this paper we will 
combine Hagedorn's and
Tsallis' approach. We will introduce a suitable
generalized grand canonical partition function. The
predictions following from this generalized thermodynamic model
can be evaluated analytically (at least in a certain
approximation).
We  will also take into account
the multiplicity and suggest a concrete dependence of the
non-extensitivity parameter $q$ on the energy $E$. The result will be
a concrete formula for differential cross sections of
transverse momenta $p_T$ where no free 
fitting parameters are left. Our formula turns out
to be in excellent agreement with experimental measurements in
$e^+e^-$ annihilation experiments.
This turns out to be true for the entire range of 
center of mass energies that have been
probed in experiments.
Our analytical formulas are also in good agreement with the curves obtained
from Monte Carlo simulations, thus suggesting that the complexity inherent
in these simulations is effectively reproduced by the simple
thermodynamic model considered here.

This paper is organized as follows. In section 2 we shortly review the
main results of Hagedorn's theory. In section 3 we shortly review
Tsallis' theory. In section 4
we combine both. 
We will write down the generalized grand canonical partition function,
including both fermions and bosons. In section 5 we will consider 
a large $p_T$ approximation,
which allows for
a simple analytical treatment. We will numerically show
that this approximation is a good one even for relatively
small values of the transverse momentum.
We derive the 
relevant form of the probability density and 
calculate all moments of
transverse momenta. In section 6 we investigate the (weak) dependence
of the Hagedorn temperature $T_0$ on the non-extensitivity parameter $q$.
In section 7 we analyse 
the energy dependence of the parameter $q$.
In section 8 we take into account the multiplicity and derive the final formula
for the differential cross section. Finally, in section 9 we compare 
our theoretical
results with
the experimentally measured results of the TASSO and DELPHI
collaboration. Our concluding remarks are given in section 10.

\section{Hagedorn's theory}

Hagedorn's theory \cite{hage} 
essentially predicts that the differential
cross section in a scattering experiment is given by
\begin{equation}
\frac{1}{\sigma}\frac{d\sigma}{dp_T} = c p_T \int_0^\infty dx \;
e^{-\beta \sqrt{x^2+\mu^2}} \label{here2}
\end{equation}
Here $p_T$ is the transverse momentum, $\mu=\sqrt{p_T^2+m^2}$ is
the transverse mass, the integration
variable $x$ stands for the longitudinal momentum,
$\beta=1/(kT_0)$ denotes the inverse
Hagedorn temperature, and $c$ is some constant.
The factor $e^{-\beta\sqrt{x^2+\mu^2}}$ is immediately recognized as
a Boltzmann factor with energy given by the 
relativistic energy-momentum relation. In the following, we
will set the Boltzmann constant $k$ equal to 1.

The Hagedorn temperature $T_0$ is independent of the center of mass
energy $E$ of the beam and typically has a a value of about 100-300 MeV.
The physical idea underlying
Hagedorn's approach is that $T_0$ describes a kind of 'boiling temperature' 
of nuclear matter,
which cannot be further increased by external energy transfer. 
Any further increase of energy just produces new states of particles
rather than an increase of temperature.

The integral in eq.~(\ref{here2}) can be further
evaluated to yield 
\begin{equation}
\frac{1}{\sigma}\frac{d\sigma}{dp_T} =
cp_T\mu K_1(\beta \mu), 
\end{equation}
where $K_1$ is the modified
Hankel function. For $p_T$ large compared to both $T_0$ and $m$ one obtains
from the asymptotic behaviour of the Hankel function
the approximate formula
\begin{equation}
\frac{1}{\sigma}\frac{d\sigma}{dp_T} \sim p_T^{3/2} e^{-\beta p_T}.\label{y}
\end{equation}

% move away************************************************
%It is obvious that the main contribution of the integrand
%in eq.~(\ref{here2}) is for 
%small values of $x$, hence one usually works with
%the approximation of small $p_L$. Moreover, the contribution
%of the mass term $\mu^2$
%is usually neglected. Using $x^2<<u^2$ 
%we obtain 
%$\sqrt{x^2+u^2}\approx  u(1+x^2/(2u^2))$, which yields
%\begin{equation}
%\int_0^\infty dx \; e^{-\sqrt{x^2+u^2}} \approx
%\int_0^\infty dx \; e^{-u-\frac{1}{2u}x^2} =\sqrt{\frac{\pi}{2}}u^{1/2}e^{-u}.
%\end{equation}
%This means, quite a good approximation of Hagedorn's result is the formula
%\begin{equation}
%\frac{1}{\sigma} \frac{d\sigma}{dp_T} \sim u^{3/2}e^{-u}.
%\end{equation}
% **************************************************************

One sees that the 
differential cross section
decays exponentially for large values of $p_T$.
While this exponential decay is indeed observed for collider
experiments with relatively small center
of mass energies ($E< 10$ GeV), clear deviations from
exponential decay have been
observed at higher energies \cite{TASSO, DELPHI, ZEUS}. 
Here the Hagedorn theory is not valid any more and the dependence
of the differential cross section on the transverse momentum
is more complicated.
Indeed, for large $p_T$ polynomial decay is observed.
For example,
the ZEUS colloboration \cite{ZEUS} has fitted their measurements 
obtained in $ep$ collision experiments
by an empirical power law of the form $(1+const \cdot p_T)^{-\alpha}$, with
exponent $\alpha$ measured as $\alpha = 5.8 \pm 0.5$.

In the following sections we will
show that it is straightforward to extend the
Hagedorn theory in such a way that it decribes experiments
at high energies ($E>10$ GeV)
as well. The basic tool for this is the  
recently developed formalism of non-extensive statistical mechanics.
It will just lead to asymptotic power laws of the above form.

\section{Tsallis' theory}

The formalism of non-extensive statistical mechanics
is a generalization of the
ordinary formalism of statistical
mechanics \cite{1}--\cite{3}. Wheras ordinary statistical mechanics is
derived by extremizing
the Boltzmann-Gibbs entropy $S= -\sum_i p_i \ln p_i$
(subject to constraints), in non-extensive 
statistical mechanics
the more general Tsallis entropies
\begin{equation}
S_q= \frac{1}{q-1} \left( 1- \sum_i p_i^q \right)
\end{equation}
are extremized. 
The $p_i$ are probabilities associated with the microstates of a
physical system, and
$q$ is the non-extensivity parameter. The
ordinary Boltzmann-Gibbs entropy is obtained in the limit
$q \to 1$.

The Tsallis entropies are closely related
to the R\'{e}nyi information measures \cite{renyi}
and have similarly nice properties as the
Boltzmann-Gibbs entropy has. They are positive, concave,
take on their extremum for the uniform distribution, and
preserve the Legendre transform structure of thermodynamics.
However,
they are not additive for independent
subsystems (hence the name 'non-extensive' statistical mechanics).
In the mean time
a lot of physical applications have been reported
for the formalism with $q\not= 1$.
Examples are
3-dimensional fully
developed hydrodynamic turbulence \cite{hydro,ari,12},
2-dimensional turbulence in pure 
electron plasmas \cite{9a,9}, 
Hamiltonian systems with long-range interactions \cite{6}, granular
systems \cite{7}, systems with strange non-chaotic attractors \cite{8},
and peculiar velocities in Sc galaxies \cite{10}.

Given some set of probabilities $p_i$ one can proceed to another set
of probabilities $P_i$ defined by
\begin{equation}
P_i = \frac{p_i^q}{\sum_i p_i^q}.
\end{equation}
These distributions are called escort distributions \cite{5}.
Extremizing $S_q$ under the
energy constraint
\begin{equation}
\sum_i P_i \epsilon_i =U_q,
\end{equation}
where the $\epsilon_i$ are the energy levels of the microstates,
the probabilities $P_i$ come out of the extremization procedure as
\begin{equation}
P_i= \frac{1}{Z_q} (1+(q-1)\beta \epsilon_i )^{-\frac{q}{q-1}},
\end{equation}
where
\begin{equation}
Z_q= \sum_i (1+(q-1) \beta \epsilon_i)^{-\frac{q}{q-1}}
\end{equation}
is the partition function and $\beta=1/T$ is a suitable
inverse temperature variable (depending on $U_q$). 
In the limit $q \to 1$, ordinary
statistical mechanics is recovered, and the $P_i$ just reduce
to the ordinary
canonical distributions $P_i=\frac{1}{Z} e^{-\beta \epsilon_i}$.
For $q\not=1$, on the other hand,
they can be regarded as generalized versions of the
canonical ensemble, describing probability distributions
in a complex non-extensive system at inverse temperature $\beta$.

\section{Combining Hagedorn's and Tsallis' theory}

\subsection{States of factorizing probabilities}

To generalize Hagedorn's theory we first have to decide
on how to introduce grand canonical partition functions
in non-extensive statistical mechanics. The problem is non-trivial,
as one immediately recognizes from just considering a
system of $N$ independent particles. 
In fact, it must first
be defined what one means by independence in the non-extensive
approach. The most plausible
definition is that for independent particles probabilities should
factorize. Let us consider such a distinguished factorized
state where each joint probability is given by products of 
single-particle Tsallis distributions.
This means, the joint probability to $P_{i_1, i_2, \ldots ,i_N}$
to observe particle 1 in energy state $\epsilon_{i_1}$,
particle 2 in energy state $\epsilon_{i_2}$, and so on
is given by
\begin{equation}
P_{i_1, i_2, \ldots , i_N} =\frac{1}{Z} \prod_{j=1}^N(1+ (q-1)\beta 
\epsilon_{i_j})^{-\frac{q}{q-1}}. \label{e1}
\end{equation}
At the same time, we could also describe our system by the
Tsallis distribution formed with the total energy 
(the Hamiltonian $H(i_1,i_2,\ldots,i_N)$
of the system)
\begin{equation}
P_{i_1, i_2, \ldots ,i_N} =\frac{1}{Z}
(1+ (q-1)\beta H(i_1,\ldots ,i_N))^{-\frac{q}{q-1}}.     \label{e2}
\end{equation}
Equating (\ref{e1}) and (\ref{e2}) 
we see that the total energy of the system
is given by
\begin{equation}
1+(q-1)\beta H=\prod_{j=1}^N (1+(q-1)\beta\epsilon_{i_j}),
\end{equation}
which we may write as
\begin{equation}
H=\sum_j \epsilon_{i_j} +(q-1)\beta \sum_{j,k}\epsilon_{i_j}\epsilon_{i_k}
+(q-1)^2\beta^2 \sum_{j,k,l} \epsilon_{i_j}\epsilon_{i_k}\epsilon_{i_l}+\cdots
\label{inter}
\end{equation}
(all summation indices are pairwise different).
This means that for the unique particle state where probabilities factorize
the total energy of the system is not the sum of the single particle energies,
provided $q\not= 1$. In other words, if seen from the
energy point of view, the particles are formally interacting with a
coupling constant $(q-1)\beta$
although the probabilities factorize. This fact is not too surprising---
the formalism of non-extensive statistical mechanics is of course designed to
describe systems with (long-range) interactions, and also the
entropy is non-additive.

We can also invert
the above statement. If we consider a Hamiltonian that is just the
sum of the single particle energies, then the probabilities do not
factorize provided $q\not= 1$. This is a well-known statement
in non-extensive statistical mechanics. 

In the following, we will generalize Hagedorn's theory using the unique
states of factorizing probabilities---thus implicitly introducing
interactions between the particles from the energy point of view.

\subsection{Statistical mechanics of the fireball}

We will consider particles of different types and label the particle types
by an index $j$. Each particle can be in a certain momentum state labelled
by the index $i$.
The energy associated with this state is
\begin{equation}
\epsilon_{ij}=\sqrt{{\bf p}^2_i+m_j^2}
\end{equation}
(the relativistic energy-momentum relation).
Let us define a non-extensive Boltzmann factor by
\begin{equation}
x_{ij}=(1+(q-1)\beta\epsilon_{ij})^{-\frac{q}{q-1}}.
\end{equation}
It approaches the ordinary Boltzmann factor $e^{-\beta\epsilon_{ij}}$
for $q\to 1$. We now very much follow
Hagedorn's original ideas, replacing the ordinary Boltzmann factor by
the generalized one.
The generalized grand
canonical partition function is introduced as
\begin{equation}
Z=\sum_{(\nu)} \prod_{ij}x_{ij}^{\nu_{ij}}
\end{equation}
Here $\nu_{ij}$ denotes the number of particles of type $j$ in
momentum state $i$. The sum
$\sum_{(\nu)}$ stands for a summation over all possible particle numbers.
For bosons one has $\nu_{ij}=0,1,2,\ldots ,\infty$,
whereas for fermions one has $\nu_{ij}=0,1$. It follows that
for bosons 
\begin{equation}
\sum_{\nu_{ij}} x_{ij}^{\nu_{ij}}=\frac{1}{1-x_{ij}} \;\;\;\;\;\;\; (bosons)
\end{equation}
whereas for fermions
\begin{equation}
\sum_{\nu_{ij}} x_{ij}^{\nu_{ij}}=1+x_{ij} \;\;\;\;\;\;\;(fermions)
\end{equation}
Hence the partition function can be written as
\begin{equation}
Z=\prod_{ij}\frac{1}{1-x_{ij}}\prod_{i'j'}(1+x_{i'j'}),
\end{equation}
where the prime labels fermionic particles. For the logarithm
of the partition function we obtain
\begin{equation}
\log Z= -\sum_{ij} \log(1-x_{ij})+\sum_{i'j'}\log(1+x_{i'j'}).
\end{equation}
One may actually proceed to continuous variables 
by replacing
\begin{equation}
\sum_i [...] \to \int_0^\infty \frac{V_0 4 \pi p^2}{h^3}[...]dp
=\frac{V_0}{2\pi^2} \int_0^\infty p^2[...]dp \;\;\;\;(\hbar=1)
\end{equation}
($V_0$: volume of the interaction region) and
\begin{equation}
\sum_j [...] \to \int_0^\infty \rho(m)[...]dm,
\end{equation}
where $\rho(m)$ is the mass spectrum. 

Let us now calculate the average occupation number of a 
particle of species $j$ in the momentum state $i$. We obtain
\begin{eqnarray}
\bar{\nu_{ij}} =x_{ij} \frac{\partial}{\partial x_{ij}} \log Z
&=& \frac{x_{ij}}{1\pm x_{ij}} \nonumber \\
&=& \frac{1}{(1+(q-1)\beta\epsilon_{ij})^{\frac{q}{q-1}}\pm 1}
\end{eqnarray}
where the $-$ sign is for bosons and the $+$ sign for fermions.

In order to single out a particular particle of mass $m_0$,
one can formally work with the mass spectrum $\rho(m)=\delta(m-m_0)$.
To obain the probability to observe a particle of mass $m_0$ in
a certain momentum state, we have to multiply
the average occupation number with the available volume in momentum
space. An infinitesimal volume in momentum space can be written as
\begin{equation}
dp_xdp_ydp_z=dp_Lp_T\sin\theta dp_T d\theta
\end{equation} where $p_T=\sqrt{p_y^2+p_z^2}$ is the transverse monentum and 
$p_x=p_L$ is the longitutinal one.  By integrating over all $\theta$ and $p_L$ 
one finally arrives at a 
probability density $w(p_T)$ of transverse momenta given by \begin{equation}
w(p_T) =const \cdot p_T \int_0^\infty dp_L 
\frac{1}{(1+(q-1)\beta\sqrt{p_T^2+p_L^2+m_0^2})^{\frac{q}{q-1}}\pm 1}.
\end{equation}
Since the Hagedorn temperature is rather small (of the order of the $\pi$ mass),
under normal circumstances one has $\beta \sqrt{p_T^2+p_L^2+m_0^2} >>1$,
and hence the $\pm 1$ can be neglected if $q$ is close
to 1. One thus obtains for both
fermions and bosons the statistics
\begin{equation}
w(p_T)\approx const \cdot p_T \int_0^\infty dp_L 
(1+(q-1)\beta\sqrt{p_T^2+p_L^2+m_0^2})^{-\frac{q}{q-1}}
\end{equation}
which, if our theory is correct,
should determine the $p_T$ dependence of experimentally
measured particle spectra. The differential
cross section $\sigma^{-1} d\sigma /dp_T$ is expected to be
proportional to $w(p_T)$.

\section{Large $p_T$ approximation}

\subsection{The differential cross section}

To further proceed with analytic calculations, one may actually perform
one further approximation step. This is the generalization of the step
of 
going from eq.~(\ref{here2}) to eq.~(\ref{y}) in Hagedorn's original
theory, which is a good approximation for large $p_T$.
Since for our applications $q$ is close to 1, one
expects that a similar step is possible in the more general non-extensive
theory.

Let us write the formula for the differential cross section
in the form
\begin{equation}
\frac{1}{\sigma} \frac{d\sigma}{dp_T} = cu \int_0^\infty
dx \; \left( 1+(q-1)\sqrt{x^2+u^2+m_\beta^2} \right)^{-\frac{q}{q-1}}.
\label{exact} \end{equation}
Here $x=p_L/T_0$, $u=p_T/T_0$ and $m_\beta:=m_0/T_0$ are the
longitudinal momentum, transverse momentum and mass in units of
the Hagedorn temperature $T_0$, respectively. $c$ is a suitable
constant. Let us
look at this formula for large values of $p_T$. If $u$
is very large, we can approximate $\sqrt{x^2+u^2+m_\beta^2}= u\sqrt{1+(x^2+m_\beta^2)/u^2}
\approx u+(x^2+m_\beta^2)/(2u)$. Of course, for this to be true the
integration variable $x$ should not be too large, but for large $x$
the integrand is small anyway and yields only a very small contribution
to the cross section.
Moreover, since $u$ is large we may 
also neglect the mass term $m_\beta^2$, arriving at
the aprroximation
\begin{equation}
\frac{1}{\sigma} \frac{d\sigma}{dp_T} \approx cu \int_0^\infty
dx \; \left( 1+(q-1)\left( u+\frac{x^2}{2u} \right) \right)^{-\frac{q}{q-1}}
\label{appro}
\end{equation}
Although this may look like a rather crude approximation, in practice it is
quite a good one. This is illustrated in Fig.~1, which shows
the right hand sides
of eq.~(\ref{exact}) and eq.~(\ref{appro}) for $q=1,1.1,1.2$. The lines
corresponding to the exact expression (\ref{exact})
and the approximation (\ref{appro}) can hardly be distinguished.
The range of $u$ and $q$ 
is similar to what we will use
later for the comparison with experimental measurements. Since
the Hagedorn temperature $T_0$ is about 120 MeV, and since very often
pions are produced, we have chosen in
formula (\ref{exact}) $m_\beta^2=(m_{\pi}/T_0)^2\approx 1.3$.
The errors made in eq.~(\ref{appro})
by neglecting $m_\beta$ and 
by approximating the square root work in opposite directions and
almost cancel each other for $q\approx 1.1$.

Within this approximation we can now easily proceed by analytical means.
We may write eq.~(\ref{appro}) as
\begin{equation}
\frac{1}{\sigma} \frac{d\sigma}{dp_T}
= cu \left( 1+(q-1)u \right)^{-\frac{q}{q-1}}
\int_0^\infty dx \left( 1 +\frac{q-1}{2u(1+(q-1)u)}x^2\right)^{-\frac{q}{q-1}}
\end{equation}
Substituting
\begin{equation} 
t:= \sqrt{\frac{q-1}{2u(1+(q-1)u)}}\; x 
\end{equation}
and
using the general formula \cite{13}
\begin{equation}
\int_0^\infty \frac{t^{2x-1}}{(1+t^2)^{x+y}}dt =\frac{1}{2}B(x,y)
\;\;\;\;\;\;(Re \; x>0, Re \; y >0), \label{here33}
\end{equation}
where
\begin{equation}
B(x,y) =\frac{ \Gamma (x) \Gamma (y) }{\Gamma (x+y)}
\end{equation}
denotes the beta-function,
one arrives at
\begin{equation}
\frac{1}{\sigma} \frac{d\sigma}{dp_T} =c(2(q-1))^{-1/2}
B\left( \frac{1}{2}, \frac{q}{q-1}-\frac{1}{2} \right) u^{3/2} \left(
1+(q-1)u\right)^{-\frac{q}{q-1}+\frac{1}{2}}.
\end{equation}
This formula, with suitably determined $c$, $q$, $T_0$, will turn out to be
in very good agreement with experimentally measured
cross sections.

\subsection{Normalization}
Let us quite generally consider a probability density with the above 
$u$-depen\-dence
\begin{equation}
p(u)=\frac{1}{Z_q} \; 
u^{3/2} \left( 1+(q-1)u \right)^{-\frac{q}{q-1}+\frac{1}{2}}
\label{here3}
\end{equation}
Normalization yields 
\begin{equation}
Z_q=\int_0^\infty u^{3/2} \left( 1+(q-1)u\right)^{-\frac{q}{q-1}
+\frac{1}{2}}du.
\end{equation}
Substituting $t^2:=(q-1)u$ the integral can be brought
into the form (\ref{here33}), and one obtains 
\begin{equation}
Z_q=(q-1)^{-5/2} B\left( \frac{5}{2}, \frac{q}{q-1}-3 \right)
\end{equation}
Note that the beta function further simplifies if $\frac{q}{q-1}-3$ is an
integer. For
$n \in {\bf N}$ one generally has 
\begin{equation}
B(x,n) = \frac{(n-1)!}{x(x+1)(x+2)\cdots(x+n-1)},
\end{equation}
in our case $x=\frac{5}{2}$.

\subsection{Moments}
Let us now evaluate the moments of $u$ defined by
\begin{equation}
\langle u^m \rangle =\int_0^\infty u^m p(u) du 
=\frac{1}{Z_q}\int_0^\infty u^{\frac{3}{2}+m}\left( 1+(q-1)u\right)^{-\frac{q}{q-1}
+\frac{1}{2}}du.
\end{equation}
Again substituting $t^2=(q-1)u$ the integral can be evaluated to give
\begin{eqnarray}
\langle u^m \rangle &=&\frac{1}{(q-1)^m}\frac{B\left( \frac{5}{2}+m,
\frac{q}{q-1}-3-m \right)}{B\left( \frac{5}{2}, \frac{q}{q-1}-3 \right)}
\nonumber \\
&=& \frac{1}{(q-1)^m} \frac{\Gamma \left( \frac{5}{2}+m \right)
\Gamma \left( \frac{q}{q-1}-3-m \right)}{\Gamma \left( \frac{5}{2} \right)
\Gamma \left( \frac{q}{q-1}-3 \right)} \label{here4}
\end{eqnarray}
Generally the Gamma function satisfies
\begin{equation}
\Gamma (x+1)=x\Gamma (x)
\end{equation}
which one may iterate to obtain
\begin{equation}
\Gamma (x+m) = \Gamma (x) \prod_{j=0}^{m-1} (x+j).
\end{equation}
Using this in eq.~(\ref{here4}) one finally arrives at
\begin{equation}
\langle u^m \rangle = \frac{1}{2^m} \prod_{j=0}^{m-1} \frac{5+2j}{4+j-(3+j)q}.
\end{equation}
In particular, one obtains for the average of $u$ 
\begin{equation}
\langle u \rangle = \frac{1}{2} \frac{5}{4-3q}
\end{equation}
and for the second and third moment
\begin{eqnarray}
\langle u^2 \rangle &=& \frac{1}{4} \frac{5}{4-3q} \frac{7}{5-4q} \\
\langle u^3 \rangle &=& \frac{1}{8} \frac{5}{4-3q} \frac{7}{5-4q}
\frac{9}{6-5q}.
\end{eqnarray}
The variance is given by
\begin{equation}
\sigma^2:= \langle u^2\rangle -\langle u \rangle^2
=\frac{5}{4} \frac{3-q}{(4-3q)^2(5-4q)}.
\end{equation}
Note that the moments obey the simple recurrence relation
\begin{equation}
\langle u^{m+1} \rangle = \langle u^m \rangle \cdot \frac{5+2m}{4+m-(3+m)q}.
\end{equation}

\section{$q$-dependence of the Hagedorn temperature}

A fundamental property of Hagedorn's theory is the fact that
the Hagedorn temperature $T_0$ is independent of the center of
mass energy $E$.
Now, in the generalized theory we have a new parameter $q$
and it is a priori not clear if and how $T_0$ depends on $q$.

However, looking at eq.~(\ref{inter}) we recognize that 
if $q$ increases from 1 to slightly larger values the (formal)
interaction energy of our system increases. 
This increase in interaction energy must come from somewhere
and is expected to be taken from the (finite volume) heat bath of the fireball.
Thus the effective temperature of the bath is expected
to slightly decrease with increasing $q$.

To get a rough estimate of this effect, let us 
work within the approximation scheme of the previous section.
Differentiating eq.~(\ref{here3}) with respect to $u$
one immediately sees that the
distribition $p(u)$ has a maximum at
\begin{equation}
u^*=\frac{p_T^*}{T_0}=\frac{3}{3-q}. \label{tq}
\end{equation}
On the other hand, the experimentally measured cross sections
always appear to have their maximum at roughly the same value of
the transverse 
momentum $p_T$, namely at
\begin{equation}
p_T^* \approx 180 \; MeV,
\end{equation}
independent of the beam energy $E$. This implies
that in the generalized thermodynamic approach the effective
Hagedorn temperature $T_0$ will become (slightly) $q$-dependent. 
\begin{equation}
T_0=\left( 1-\frac{q}{3} \right) p_T^* \label{tqq}
\end{equation}
The variation of $T_0$ with energy $E$ is actually very weak. In 
the next section we will consider two extreme cases, namely
$q\to 1$ for $E\to 0$ and $q\to \frac{11}{9}$ for $E\to \infty$.
The first case corresponds to $T_0=120$ MeV, the second case
to $T_0=107$ MeV. This is only a very small variation.

\section{Energy dependence of $q$}

We still have to decide how the non-extensitivity parameter $q$
depends on the center of mass energy $E$ of the beam. We will present some
theoretical arguments, which are well 
supported by the experimental data.

\subsection{Plausible value for $q_{max}$}

Let us define $\alpha$ to be the power of the term
$(1+(q-1)u)^{-1}$ in eq.~(\ref{here3})
\begin{equation}
\alpha =\frac{q}{q-1} -\frac{1}{2}. \label{here5}
\end{equation}
Clearly, for small energies $E$ Hagedorn's theory ($q=1$) is
valid:
\begin{equation}
E \to 0 \Longrightarrow \alpha \to \infty \Longleftrightarrow q\to 1
\end{equation}
For increasing $E$, $\alpha$ should decrease. For example,
the ZEUS collaboration measures $\alpha =5.8 \pm 0.5$
at medium energies \cite{ZEUS}. However, $\alpha$ cannot
become arbitrarily small, because we must have a finite variance
of $u$ (otherwise statistical mechanics does not make sense). For
large $u$, we can certainly use the large $p_T$
approximation of section 5. Asymptotically the density $p(u)$ decays as
\begin{equation}
p(u) \sim u^{-\alpha+3/2}
\end{equation}
Thus $u^2p(u)$ decays as $u^{-\alpha+7/2}$ and hence $\langle u^2 \rangle
=\int du \; u^2p(u)$ only exists if
\begin{equation}
-\alpha +\frac{7}{2} < -1 \Longleftrightarrow \alpha > \frac{9}{2}
\Longleftrightarrow q < \frac{5}{4}
\end{equation}
(for $\alpha =9/2$ there is a logarithmic divergence). 
This is a strict upper bound on $q$.

If, in addition,
we postulate that the distinguished limit
theory corresponding to the largest possible value
of $q$ should be analytic in $\sqrt{u}$,
then only integer and half-integer values of $\alpha$ are allowed.
This leads to
\begin{equation}
\alpha_{min}=5 \Longleftrightarrow q_{max}=\frac{11}{9}=1.222
\end{equation}
as the smallest possible value of $\alpha$, respectively largest
value of $q$.

\subsection{A smooth dependence on $q$}

It is reasonable to assume (and supported by
the experimental results) that the smallest possible value of
the exponent $\alpha$ is taken on for the largest possible
energy $E$. Hence we have the two limit cases
\begin{eqnarray}
E\to 0 &\Longrightarrow& \alpha \to \infty \\
E\to \infty &\Longrightarrow& \alpha \to \alpha_{min}=5.
\end{eqnarray}
A smooth monotonously decreasing interpolation between these
limit cases is given by the formula
\begin{equation}
\alpha (E) =\frac{5}{1-e^{-E/E_0}}, \label{here6}
\end{equation}
where $E_0$ is some suitable energy scale. It turns out that the experimental
data are perfectly fitted if we choose
\begin{equation}
E_0=\frac{1}{2}m_Z =45.6 \; GeV,
\end{equation}
$m_Z$ being the mass of the $Z^0$ boson.

Whether this coincidence of the relaxation parameter $E_0$
with half of the $Z^0$ mass
is a random effect or
whether there is physical meaning behind this 
is still an open question. What is clear is the following.
At center of mass energies of the order 90 GeV
often a massive elektroweak gauge boson $Z^0$ is produced.
If a $Z^0$ at rest decays into
a quark pair $q\bar{q}$, then each quark has the initial
energy $\frac{1}{2}m_Z$, which is input to the hadronization
cascade. 
It is obvious that at this energy scale there is a rapid change
in the behaviour of the fireball due to the occurence of the
electroweak bosons.
The same mass scale now seems to mark the crossover scale between
ordinary Hagedorn thermostatistics, valid for $E <<\frac{1}{2}m_Z$,
and generalized thermostatistics with $q \approx q_{max}$, valid
for $E>>\frac{1}{2}m_Z$. 

Eq.~(\ref{here6}) can equivalently be written as
\begin{equation}
q(E)=\frac{11-e^{-E/E_0}}{9+e^{-E/E_0}}. \label{here10}
\end{equation}

\section{Energy dependence of the multiplicity}

The differential cross section $\sigma^{-1} d\sigma/dp_T$ measured
in experiments is not a normalized probability density. This is
due to the fact that it is dependent on the multiplicity $M$
(the average number of produced charged particles). Also, it has 
dimension $GeV^{-1}$, whereas the probability density $p(u)$ is
dimensionless. All this, however, is just a question
of normalization.

To proceed from $p(u)$ to $\sigma^{-1} d\sigma/dp_T$, we should multiply
$p(u)$ with the multiplicity $M$. Also, in order to give the correct
dimension of a cross section, we should multiply with $T_0^{-1}$.
Thus we arrive at the formula
\begin{equation}
\frac{1}{\sigma}\frac{d\sigma}{dp_T}= \frac{1}{T_0} M p(u) \label{here7}.
\end{equation}
The multiplicity $M$ as a function of the beam energy $E$ has been independently
measured in many experiments. A good fit of the 
experimental data in the relevant energy region
is the formula 
\begin{equation}
M=\left( \frac{E}{T_0^{q=1}}\right)^{5/11} \;\;\;\;T_0^{q=1}=120 \; MeV 
\label{here88}
\end{equation}
(see Fig.~2, data as collected in \cite{passon}). 
It is remarkable that the constant in front of this power law is 1
if $E$ is measured in units of the Hagedorn Temperature $T_0^{q=1}$.
Apparently, the Hagedorn temperature 
is a very appropriate portion of energy for our statistical
approach.
Essentially, the scaling law (\ref{here88}) says that 
only a certain fraction of the logarithm of the energy $E$ (described by the
scaling exponent $5/11\approx 0.45$) is used to increase
the number of charged particles.

\section{Comparison with experimentally measured cross sections}

We are now in a position to directly compare with experimental
measurements of cross sections. All parameters of the
generalized Hagedorn theory 
(the Hagedorn temperature $T_0(q)$,
the non-extensitivity parameter $q(E)$ and the normalization constant of
the cross section) have been discussed in the previous sections
and concrete equations have been derived. 

Formula (\ref{here7}) with $p(u)$ given by
eq.(\ref{here3}), $q(E)$ given by eq.~(\ref{here10}),
$T_0(q)$ given by eq.(\ref{tqq})
and multiplicity $M(E)$ given by eq.~(\ref{here88})
turns out to very well reproduce
the experimental results of cross sections
for all energies $E$. This is illustrated
in Fig.~3, which shows experimental cross sections
measured by the
TASSO and DELPHI collaborations (\cite{TASSO, DELPHI},
data as collected in \cite{passon})
as well as our theoretical prediction. For all curves we have chosen
the same universal parameters
$p_T^*=180$ MeV and $E_0=\frac{1}{2}m_Z$, so we do not use any 
energy-dependent fitting 
parameters.

The agreement is remarkably good. Hence the statistical approach presented
in this paper qualifies as a simple thermodynamic model that explains the
experimental data quite well. In particular,
for the first time analytical
formulas are obtained that 
correctly describe the measured cross sections. The quality of
agreement is at least as good as that of Monte Carlo simulations,
in spite of the fact
that Monte Carlo simulations usually use a large
number of free parameters.
For small $p_T$, the agreement seems even to be slightly better than 
that of Monte Carlo simulations (see
e.g.\ \cite{passon} for typical Monte Carlo results). 

Our approach yields $q$-values of similar order of magnitude
as the ones used in the fits in \cite{11}. 
In our theoretical approach all
parameters are now given by concrete formulas. 
Some of these formulas,
most importantly
eq.~(\ref{here10}), have been derived empirically and 
should thus be further checked and
possibly refined by further experimental measurements.

\section{Conclusion}

In this paper we have developed a thermodynamic model
describing the statistics of transverse momenta of particles
produced in high-energy collisions. Our approach is based on
a generalization of Hagedorn's theory using Tsallis' formalism
of non-extensive statistical mechanics. Hagedorn's original
theory is recovered for small center of mass energies, whereas
for larger energies deviations from ordinary Boltzmann-Gibbs statistics 
become relevant. The crossover energy scale between the two regimes 
$E \to 0$, $q\approx 1$ and $E\to \infty$, $q\approx \frac{11}{9}$
is
given by half of the $Z^0$ mass. 

At large energies
the generalized thermodynamic theory implies that cross sections 
decay with a power law. This power law is indeed
observed in experiments. Our theory allows us 
to analytically evaluate
moments of arbitrary order  
(within a certain approximation).
We obtain 
formulas for differential cross sections that are in very good agreement
with experimentally measured cross sections in annihilation experiments.

We suggest that
future analysis of the experimental data should concentrate on precision
measurements of the
energy dependence of the parameter $q$. For this the mass spectrum
of produced particles should carefully be taken into account.
Generally,
it appears
that high-energy collider experiments do not only yield valuable
information on particle physics, but they may also
be regarded as ideal test grounds to verify new ideas from 
statistical mechanics.

\subsection*{Acknowledgement}

This research was supported in part by the National Science
Foundation under Grant No. PHY94-07194. The author gratefully
acknowledges support by a
Royal Society Leverhulme Trust Senior Research Fellowship.

\newpage

\subsection*{Figure captions}

$\,$

{\bf Fig.~1} Comparison between the exact
formula (\ref{exact}) with $m_\beta^2=(m_\pi/T_0)^2=1.3$ and 
the approximation (\ref{appro}) for
various values of $q$. The constant $c$ was set equal to 1 for all $q$.

\vspace{0.5cm}
{\bf Fig.~2} Dependence of the multiplicity $M$ on the center of mass
energy $E$. The figure shows the experimental data and a
straight line that corresponds to the scaling law (\ref{here88}). 

\vspace{0.5cm}
{\bf Fig.~3} Differential cross section as a function of the
transverse momentum $p_T$ for various center of mass energies $E$.
The data correspond to measurements of the TASSO ($E\leq 44$ GeV)
and DELPHI ($E\geq 91$ GeV)
collaboration. The solid lines are given by
the analytic formula (\ref{here7}).

\newpage
\begin{figure}{\epsfig{file=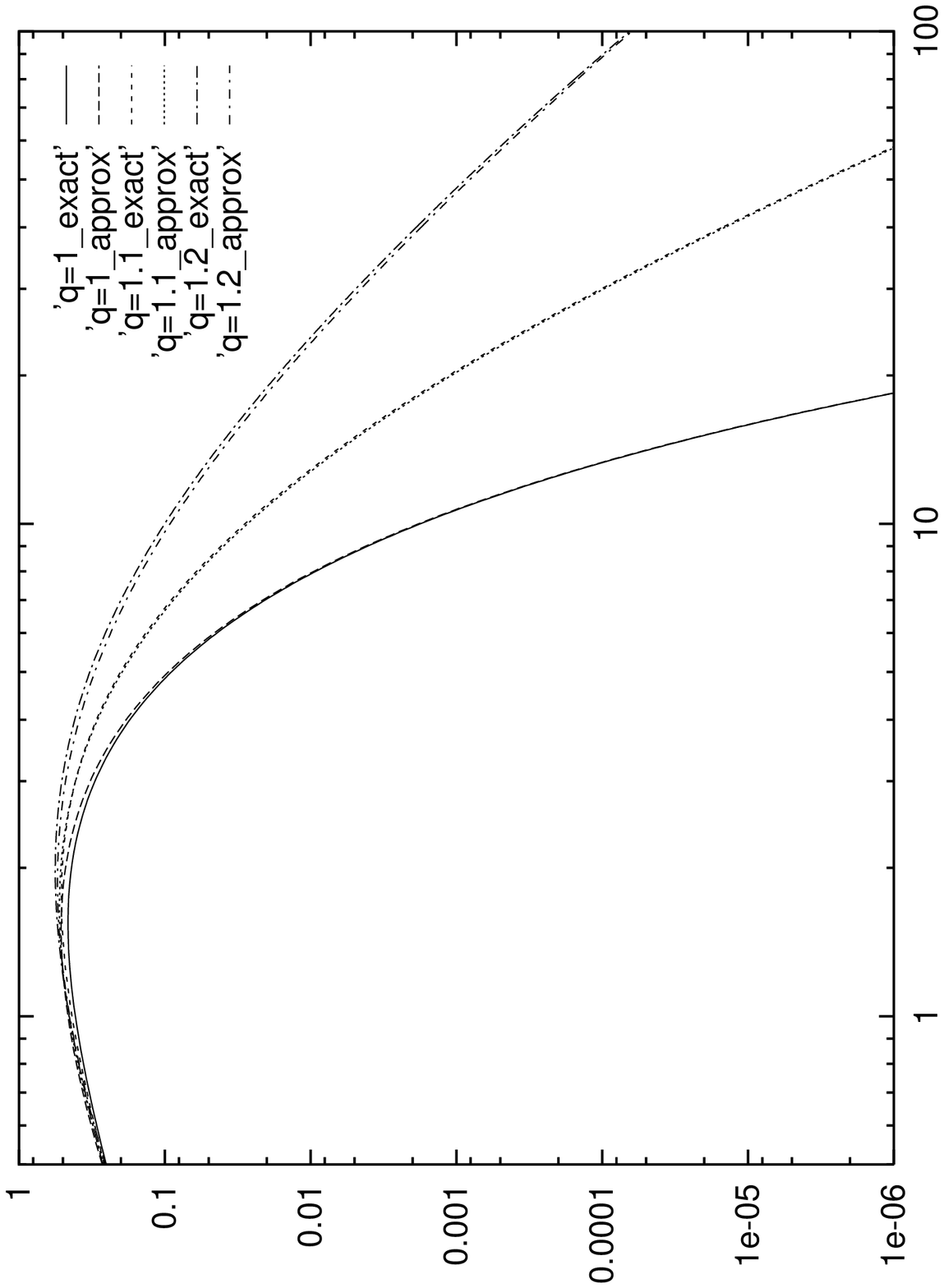, height=4.6in, width=5.2in}}
\end{figure}

\newpage
\begin{figure}{\epsfig{file=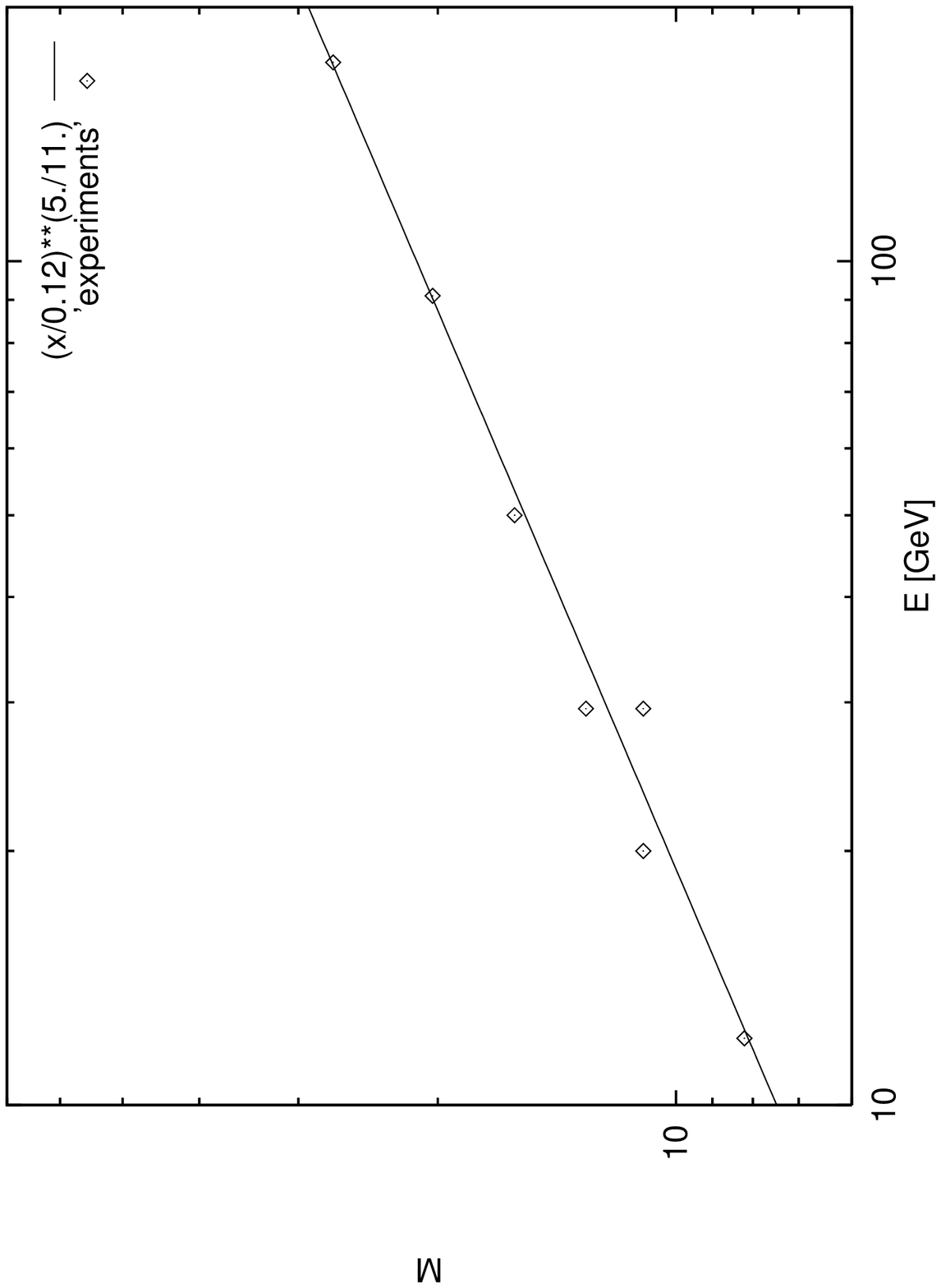, height=4.6in, width=5.2in}}
\end{figure}

\newpage
\begin{figure}{\epsfig{file=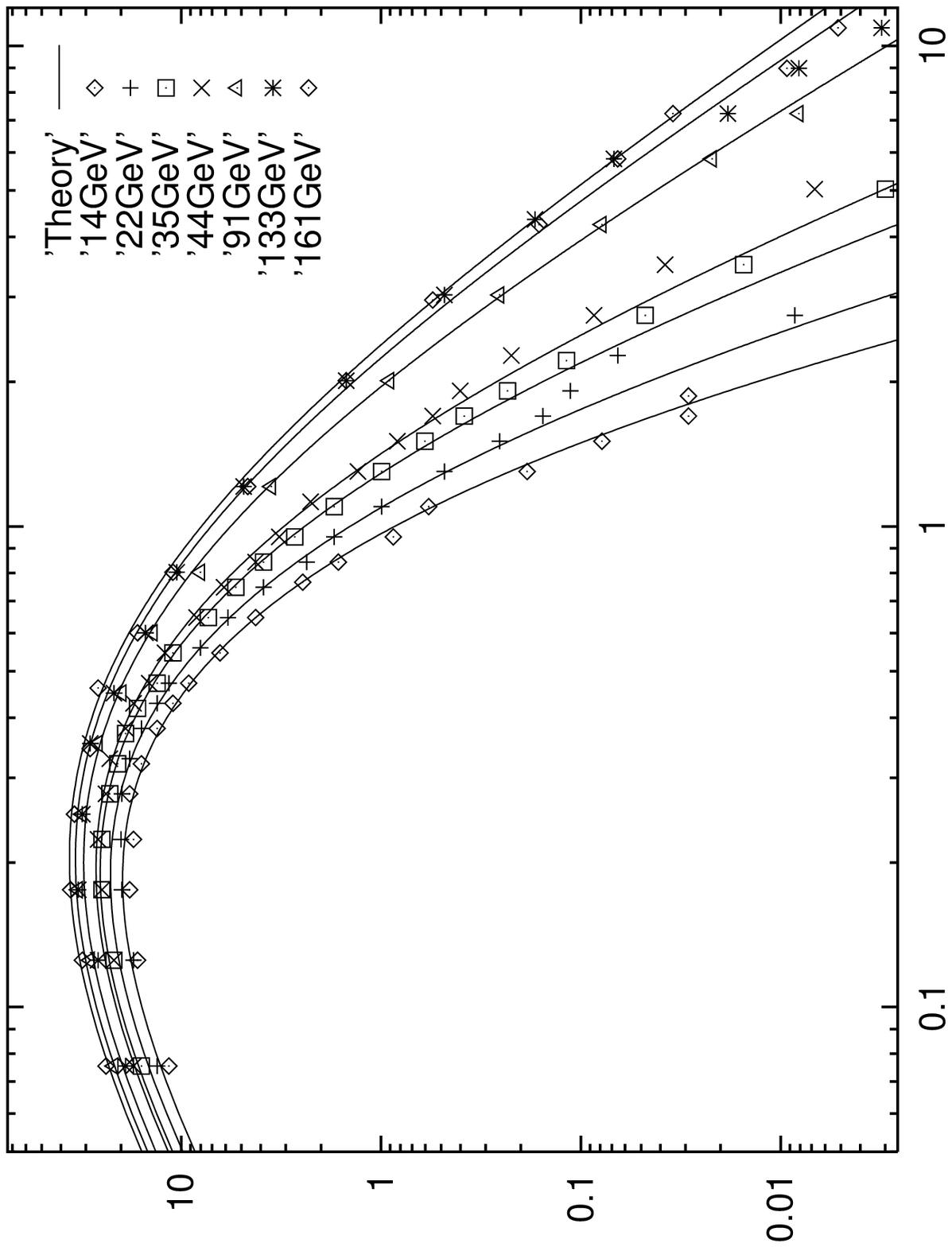, height=4.6in, width=5.2in}}
\end{figure}

\end{document}